\documentclass[a4paper,11pt]{article}
\usepackage{pos}

\usepackage{color}

\definecolor{mygreen}{RGB}{0,128,0} 
\definecolor{mybrown}{RGB}{153,102,51}

\usepackage{hyperref}
\usepackage{tikz} 
\usetikzlibrary{arrows} 
\usetikzlibrary{shapes,arrows,positioning,automata,backgrounds,calc,er,patterns}
\usepackage{tikz-feynman}
\tikzfeynmanset{compat=1.1.0}

\usepackage{color}
\usepackage{subfigure}

\usepackage{color} 
\definecolor{mygreen}{RGB}{0,128,0} 
\definecolor{mybrown}{RGB}{153,102,51}

\newcommand{\mF}{\mathcal{F}}
\newcommand{\mL}{\mathcal{L}}

\newcommand{\mN}{\mathcal{N}}
\newcommand{\mO}{\mathcal{O}}

\title{Redefining 
Higgs interactions 
at the TeV scale}

\author[a]{Rafael L. Delgado}
\author*[b]{Raquel G\'omez-Ambrosio}
\author[c]{Javier Mart\'\i nez-Mart\'\i n}
\author[d]{Alexandre Salas-Bern\'ardez}
\author[c]{Juan J. Sanz-Cillero}

\affiliation[a]{Dept. Matem\'atica Aplicadas a las TIC, Universidad Polit\'ecnica de Madrid, Nikola Tesla, s/n, 28031-Madrid, Spain}
\affiliation[b]{Dipartimento di Fisica, Universit\`a di Torino, and INFN, Sezione di Torino,
Via P. Giuria 1, 10125 Torino, Italy}
\affiliation[c]{Dept. F\'\i sica Te\'orica and IPARCOS, Universidad Complutense de Madrid, Plaza de las Ciencias 1, 28040 Madrid, Spain}
\affiliation[d]{Instituto de Física Corpuscular (IFIC), Universidad de Valencia-CSIC,
E-46980 Valencia, Spain}

\emailAdd{rafael.delgado@upm.es}
\emailAdd{raquel.gomezambrosio@unito.it}
\emailAdd{javmar21@ucm.es}
\emailAdd{asaber2@uv.es}
\emailAdd{jjsanzcillero@ucm.es}

\abstract{ We present a field redefinition 
that simplifies the Higgs Effective Field Theory 
Lagrangian for the Electroweak Symmetry Breaking Sector. 
This simplification produces the same on-shell scattering amplitudes while greatly reducing the number of contributing Feynman diagrams for $\omega\omega\to n\times h$ processes (which approximate the $W_LW_L\to n\times h$ amplitudes at the TeV scale by means of the Equivalence Theorem). 
}

\FullConference{The Eleventh Annual Conference on Large Hadron Collider Physics (LHCP2023)\\
 22-26 May 2023\\
 Belgrade, Serbia\\}

\begin{document}
\maketitle

\section{Introduction}
At the energy frontier, one of the most important aspect of physics being clarified right now is the nature of the mechanism of Electroweak Symmetry Breaking (EWSB): whether it occurs as described in the Standard Model, or whether new particles or interactions influence the global $SU(2)\times SU(2)\to SU(2)$ breaking pattern (\textit{e.g.}, \cite{Dawson:2023ebe,Buchalla:2023hqk,Arco:2023sac}), which is central to the electroweak interactions.  In the context of new physics lying well above the energy frontier, the language of Effective Field Theories (EFTs) is standard (although new possibilities are being put forward \cite{Lessa:2023tqc}). Two EFTs have been proposed to parametrize this new physics: the Standard Model EFT (SMEFT) and the Higgs EFT (HEFT), 
with a more reduced number of assumptions and a further generality 
on the nature of the EWSB in the latter~\cite{LHCHiggsCrossSectionWorkingGroup:2016ypw}. 

At the TeV scale, where the energies of the scattered particles are much higher than their masses $m_h\ll E \sim \partial$,  
the much discussed Higgs potential $V(h)$ contribution actually provides a suppressed correction to the amplitude and does not play the pivotal role
it enjoys in the SM.  The relevant leading order (LO) Lagrangian which parametrizes new physics in this regime takes the form~\cite{Delgado:2023ynh},
\begin{eqnarray}
\mL_{\rm HEFT} &=& \frac{1}{2}(\partial_\mu h)^2 + \frac{1}{2}\mF(h)\, \partial_\mu\omega^a \partial^\mu \omega^a \, +\, \mathcal{O}(\omega^4) 
\, ,
\label{eq:HEFT-Lagr}
\end{eqnarray}
where the flare function, 
\begin{equation}
\mathcal{F}(h)=1 + a_1 \frac{h}{v} + a_2 \left( \frac{h}{v} \right)^2 + a_3  \left( \frac{h}{v} \right)^3 + a_4 \left( \frac{h}{v} \right)^4 + \dots\;,
\end{equation}
plays the main role in distinguishing between the SMEFT and HEFT scenarios \cite{Gomez-Ambrosio:2022qsi}. It couples an arbitrary number of Higgs bosons, $h$, and a pair of pseudo-Goldstone Bosons, $\omega$. 

Moreover, the $\omega$ scattering approximates the corresponding $W_L$ amplitudes 
at the TeV scale by virtue of the Equivalence Theorem ~\cite{Veltman:1989ud} (EqTh).

In these proceedings we present the HEFT field redefinition in~\cite{Delgado:2023ynh} 
that eliminates at will one of the $h^n \omega \omega$ derivative vertex stemming from $\mF(h)$. In particular, the removal of the $h\omega\omega$ derivative vertex is the choice that most efficiently reduces the number of diagrams for a generic process.   

\section{HEFT simplifications through field redefinitions: understanding  \texorpdfstring{$\omega\omega\to n \times h$}{ww->nxh}}
\label{app:field-redef}

\begin{figure}[!t] 
     \centering
     \begin{subfigure}[]
         \centering
         \begin{tikzpicture}[scale=1]
     \draw[dashed] (-1,1) -- (0,0)-- (-1,-1);
     \draw[] (0,0)-- (1,1);
       \draw[] (0,0)-- (1,-1);
       \draw[] (-1.25,1) node {$\omega$};
       \draw[] (-1.25,-1) node {$\omega$};
     \draw[] (1.25,1) node {$h$};
          \draw[] (1.25,-1) node {$h$}; 
\end{tikzpicture}
     \end{subfigure}
     \qquad
     \begin{subfigure}[]
         \centering
         \begin{tikzpicture}[scale=1]
     \draw[dashed] (-1,1) -- (0,0)-- (-1,-1);
     \draw[] (0,0)-- (1,1);
     \draw[] (0,0)-- (1,0);
       \draw[] (0,0)-- (1,-1);
       \draw[] (-1.25,1) node {$\omega$};
       \draw[] (-1.25,-1) node {$\omega$};
     \draw[] (1.25,1) node {$h$};
          \draw[] (1.25,-1) node {$h$}; 
           \draw[] (1.25,0) node {$h$};
\end{tikzpicture}
     \end{subfigure}
    \qquad
     \begin{subfigure}[]
         \centering
         \begin{tikzpicture}[scale=1]
     \draw[dashed] (-1,1) -- (0,0)-- (-1,-1);
     \draw[] (0,0)-- (1,1);
     \draw[] (0,0)-- (1,0.35);
     \draw[] (0,-0)-- (1,-0.35);
       \draw[] (0,0)-- (1,-1);
       \draw[] (-1.25,1) node {$\omega$};
       \draw[] (-1.25,-1) node {$\omega$};
     \draw[] (1.25,1) node {$h$};
     \draw[] (1.25,-0.35) node {$h$};
     \draw[] (1.25,0.35) node {$h$};
          \draw[] (1.25,-1) node {$h$}; 
\end{tikzpicture}
     \end{subfigure}
     \qquad
     \begin{subfigure}[]
         \centering
         \begin{tikzpicture}[scale=1]
     \draw[dashed] (-1,1) -- (0,1)--(0,-1)-- (-1,-1);
     \draw[] (0,1)-- (1,1);
     \draw[] (0,1)-- (1,0.35);
     \draw[] (0,-1)-- (1,-0.35);
       \draw[] (0,-1)-- (1,-1);
       \draw[] (-1.25,1) node {$\omega$};
       \draw[] (-1.25,-1) node {$\omega$};
     \draw[] (1.25,1) node {$h$};
     \draw[] (1.25,-0.35) node {$h$};
     \draw[] (1.25,0.35) node {$h$};
          \draw[] (1.25,-1) node {$h$}; 
\end{tikzpicture}
     \end{subfigure}
        \caption{\small {\bf a)} Only diagram contributing to the 
        process $\omega\omega\to 2h$. {\bf b)} Only diagram contributing to the  
        process $\omega\omega\to 3h$. {\bf c-d)} Only two diagrams contributing to the  
        process $\omega\omega\to 4h$. 
        We have used the simplified Lagrangian~(\ref{eq:simpler-L}) to generate these amplitudes, so every $\omega\omega h^n$ vertex carries an $\hat{a}_n$ effective coupling. 
        Note that, in addition, one needs to consider all possible permutations for the assignment of the external particles. 
        }\label{diagrams}
\end{figure}
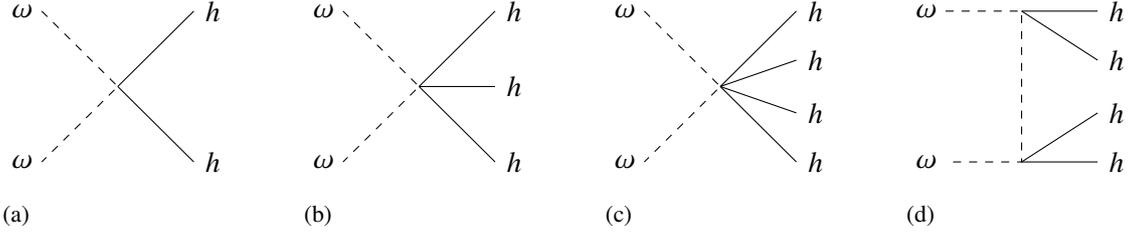

In \cite{Delgado:2023ynh}, we found that, at LO, for $2h$ and $3h$ production from 
$\omega\omega$ fusion, amplitudes are pure $s$-waves, and crossed-channel Goldstone exchanges give place to purely polynomial amplitudes,
simplifying to contact interactions. For $4h$ final states, strong cancellations also occur (see Fig. \ref{diagrams}), resulting in a contact $\omega\omega\to 4h$ interaction with one Goldstone exchange from crossed channels. We here aim to explain the origin of these cancellations. To do so, we consider field redefinitions in the form, 
\begin{eqnarray}\label{eq:C2}
&&   \omega^a\to \omega^a 
+ g(h) \, \omega^a 
\,,
\qquad    
h \to h 
+ \mathcal{N} \, (1+ g(h)) \, \omega^a\omega^a /v
\,, 
\end{eqnarray} 
with a free dimensionless real constant $\mN$, and an $\mO(h)$ function $g(h)$. In order to produce a Lagrangian with the structure of~(\ref{eq:HEFT-Lagr}), the latter is chosen to fulfill the relation $ g'(h)=- 2\mN/[ v\, \mF(h)]$, determined by the flare function $\mF(h)$ and the normalization constant $\mN$. 
This yields 
\begin{equation}
\displaystyle{    g(h)\, =\, -\, \frac{2 \mN}{v}\,  \int_0^h \frac{ds}{ \mF(s)  }     } 
\,= \mN\, \bigg( \, -2 \frac{h}{v} + 2 a \frac{h^2}{v^2} + \frac{2}{3} (b-4a^2) \frac{h^3}{v^3} + \frac{1}{2} (a_3 -4 ab  + 8 a^3) \frac{h^4}{v^4} +\mO(h^5)\bigg)  \, ,    
\label{eq:g-transform}
\end{equation} 
with the usual HEFT notation $a\equiv a_1/2$, $b\equiv a_2$. 
The application of the transformation in eq. (\ref{eq:g-transform}) to the Lagrangian~(\ref{eq:HEFT-Lagr}) leads to a new Lagrangian with exactly the same structure, but with a new function $\hat{\mF}(h)$ determined by: 
\begin{equation}
\hat{\mF}(h)\, =\, \mF(h)\, \big(1+g(h)\big)^2. 
\label{eq:Fhat}
\end{equation}

 In particular, to eliminate the linear term in $\hat{\mF}(h)$ which provide the $h\omega \omega$ vertex, we take: 
 
\begin{equation}
\mN\,=\, \frac{a}{2}\,, 
\qquad 
g(h)\,=\, -a \frac{h}{v} +a^2 \frac{h^2}{v^2} +\frac{1}{3}a(b-4 a^2) \frac{h^3}{v^3}  + \frac{1}{4} a( a_3 - 4 ab + 8 a^3) \frac{h^4}{v^4} \, +\, \mO(h^5)\, .
\label{eq:g-transform-a1=0}
\end{equation}
This choice of the scalar manifold coordinates transforms the original Lagrangian~(\ref{eq:HEFT-Lagr}) into 
\begin{eqnarray}    
\hat{\mL}_{\text{HEFT}} &=& \frac{1}{2}\partial_\mu h \partial^\mu h 
    + \frac{1}{2} \hat{\mF}(h)\, \partial_\mu\omega^a\partial^\mu\omega^a   \, \,+\, \mO(\omega^4)\, 
    \, ,  
\label{eq:simpler-L}
\end{eqnarray}      
 with the new function,  
\begin{equation}
\hat{\mF}(h)\,=\, 1 
   + \hat{a}_2 \frac{h^2}{v^2} 
   + \hat{a}_3 \frac{h^3}{v^3} 
   + \hat{a}_4 \frac{h^4}{v^4} +\,\mO(h^5) \, ,
\end{equation}
having coefficients provided by the ones of $\mF(h)$ through the combinations
\begin{eqnarray}
\hat{a}_2 &=& a_2-\frac{a_1^2}{4}\,=\, b-a^2 \, , \qquad  
\hat{a}_3 = 
a_3- \frac{2}{3}a_1 \left(a_2 -  a_1^2/4\right) \,=\,  a_3-\frac{4a}{3}\left(b-a^2\right)\,, 
\nonumber  
\\
\hat{a}_4&=& a_4 -   \frac{3}{4} a_1 a_3    +   \frac{5}{12}a_1^2\left(a_2-a_1^2/4\right) 
\,=\, a_4 -   \frac{3}{2} a\,  a_3    +   \frac{5}{3}a^2\left(b-a^2\right) \, .
\end{eqnarray}
Observe that the first significant contribution of $\hat{\mF}(h)$ occurs at $\mO(h^2)$, in contrast to $\mF(h)$, where the initial significant term emerges at $\mO(h)$. Since the generating functional of the quantum field theory is invariant under field redefinitions, both Lagrangians (\ref{eq:HEFT-Lagr}) and (\ref{eq:simpler-L}) lead to the same on-shell scattering amplitudes~\cite{Criado:2018sdb,Chisholm:1961tha}. The critical gain from this field redefinition  
is that the number of diagram topologies is greatly reduced (see Fig.~\ref{diagrams}):  
there is only 1 diagram for $\omega\omega\to 2h$ and $\omega\omega\to 3h$, and it is reduced to 2 diagram topologies for $\omega\omega\to 4h$ process (up to 
permutations in the labeling of the outgoing Higgs particles in the diagrams).


The same procedure as explained above allows us to extract the relevant combination of the flare function coefficients, $a_j$,  for a generic amplitude $\omega\omega\to n\times h$: 
\begin{enumerate}
\item Compute $g(h)$ up to $\mO(h^n)$ by plugging $\mF(h)$ in~(\ref{eq:g-transform}) up to that order with $\mN=a_1/4$ (this choice will remove $h\omega\omega$ derivative interactions).  
\item Expand $\hat{\mF}(h)=\mF(h)\,\left(1+g(h)\right)^2$ up to $h^n$. The corresponding coefficients $\hat{a}_j$ will be the relevant combinations for that process.   
Thus, following the steps above, we can easily extract the next $\hat{a}_j$ coefficients: 
$\hat{a}_5= a_5-\frac{1}{120} a_1 \left(15 a_1 \hat{a}_3-16     \hat{a}_2^2+96 \hat{a}_4  \right)$,  
$\hat{a}_6=a_6+\frac{1}{180} a_1 \left(  7 a_1 \hat{a}_2^2  -27a_1    \hat{a}_4+45\hat{a}_2 \hat{a}_3-150 \hat{a}_5 \right)$, etc.
\end{enumerate}

Data analyses that overlook this redundancy and directly fit the $a_j$ instead of the $\hat{a}_j$ is effectively introducing avoidable correlations, complicating the analysis significantly (see Fig.~2 in~\cite{Delgado:2013loa}).

In general, one can also conveniently choose the normalization $\mN$ to remove a higher order term, $a_n h^n$, from  \ $\mF(h)$ instead of the first one.  For instance, provided ${ a_2< a_1^2/4 }$,  the choice \  $\mN= \left[ a_1 \pm\sqrt{a_1^2 - 4 a_2}\right]/4$ removes the $a_2 h^2$ term in $\mF(h)$, passing this information to the terms of order $h^1$ and $h^3$, $h^4$, etc. 
Another example is provided by the normalization $\mN=\frac{3}{8}\frac{a_3}{a_2-a_1^2/4}$, which removes the $h^3$ term in $\mF(h)$ and encodes its information in the factors $\hat{a}_j$ now multiplying $h^1$, $h^2$ and $h^4$, $h^5$, etc. This detail can be important for a proper interpretation of $WW\to 3h$ computations: at high energies, in the EqTh, it is possible to describe the $\omega\omega\to 3h$ scattering without an $\omega\omega h^3$ vertex ($\hat{a}_3=0$)~\cite{Gonzalez-Lopez:2020lpd},  
understanding that the $\omega\omega h^2$ and $\omega\omega h$ couplings are not the original ones ($a_2$ and $a_1$) but some effective ones ($\hat{a}_2$ and $\hat{a}_1$).

It is interesting to note that in the dilatonic model~\cite{Halyo:1991pc,Goldberger:2007zk}, 
represented by $\mF(h)=\left(1+ a h/v\right)^2$, a significant result arises. Specifically, applying the transformation above leads to $1+ g(h)=\left(1+ a h/v\right)^{-2}$, resulting in $\hat{\mF}(h)=1$. This intriguingly leads to all $\hat{a}_j$ couplings being zero, causing all $\omega\omega\to n\times h$ amplitudes to vanish at the tree level (in the context of the EqTh). This same conclusion also applies to the Standard Model, where $a=1$.

The main drawback of this approach is that the $SU(2)_L\times SU(2)_R$ chiral invariance of the action is no longer explicit~\cite{LHCHiggsCrossSectionWorkingGroup:2016ypw}.  
The symmetry transformations become more complex in this context.  
For this reason, the correlations between the $a_1$ and $a_2$ couplings in $\mF(h)$ one finds for SMEFT-type theories with dimension $D=6$ contributions 
(this is, $a_2=2a_1-3$, found in~\cite{Gomez-Ambrosio:2022qsi,Gomez-Ambrosio:2022why})   
are no longer applicable for $\hat{\mF}(h)$ (SMEFT with $D=6$ contributions 
 does not fulfill $\hat{a}_2=2\hat{a}_1-3$ nor $\hat{a}_1=0$). The reason is that the chiral operator structures considered in~\cite{Gomez-Ambrosio:2022qsi,Gomez-Ambrosio:2022why} are deformed here for the terms of order $\omega^4$ and higher, so the conclusions therein are no longer applicable for the present simplified Lagrangian with $\hat{\mF}(h)$.      

In summary, these simplifications can be beneficial when computing electroweak processes involving Higgs and Goldstone bosons using the equivalence theorem, both at tree-level and loop-level calculations.
 
 \paragraph{Aknowledgments}
 {
This work has been supported by Spanish MICINN (PID2022-137003NB-I00,    PID2021-124473NB-I00,  PID2019-108655GB-I00/AEI/10.13039/501100011033), U. Complutense de Madrid under research group 910309, the IPARCOS institute, the EU under grant 824093 (STRONG2020), and EU COST action CA22130.  
The work of RGA is supported by the EU’s Next Generation grant DataSMEFT23 (PNRR - DM 247 08/22). The work of JMM is supported by the grant Ayudas de doctorado IPARCOS-UCM/2022. ASB acknowledges the support of the EU's Next Generation funding, grant number CNS2022-135688.
}

\bibliographystyle{JHEP}
\bibliography{references.bib}


\end{document}